\documentclass[a4paper,fleqn,usenatbib,useAMS]{mnras}

\usepackage{graphicx}	% Including figure files
\usepackage{amsmath}	% Advanced maths commands
\usepackage{amssymb}
\usepackage{multicol}        % Multi-column entries in tables
\usepackage{bm}		% Bold maths symbols, including upright Greek
\usepackage{pdflscape}	

\usepackage[T1]{fontenc}
\usepackage{ae,aecompl}

\usepackage{txfonts}

\newcommand{\be}{\begin{equation}}
\newcommand{\ee}{\end{equation}}

\title[Barred galaxies in EAGLE]{Barred galaxies in the EAGLE cosmological
  hydrodynamical simulation}

\author[Algorry et al.]{David G. Algorry$^{1,2}$\thanks{E-mail: david@oac.unc.edu.ar},
Julio F. Navarro$^{3,4}$, Mario G. Abadi$^{1,2}$, Laura V. Sales$^5$, 
 \and
Richard G. Bower$^6$, 
Robert A. Crain$^7$, 
Claudio Dalla Vecchia$^{8,9}$,
Carlos S. Frenk$^6$,
\and
Matthieu Schaller$^6$,
Joop Schaye$^{10}$, 
and Tom Theuns$^6$. \and \\
$^{1}$ Instituto de Astronom\'ia Te\'orica y Experimental, CONICET-UNC, Laprida 854, X5000BGR, C\'ordoba, Argentina\\
$^2$ Observatorio Astron\'omico de C\'ordoba, Universidad Nacional de C\'rdoba, Laprida 854, X5000BGR, C\'ordoba, Argentina\\
$^3$ Department of Physics \& Astronomy, University of Victoria, Victoria, BC V8P 5C2, Canada\\
$^4$ Senior CIfAR Fellow\\
$^5$ Department of Physics and Astronomy, University of California, Riverside, CA, 92521, USA\\
$^6$ Institute for Computational Cosmology, Department of Physics, Durham University, South Road, Durham, DH1 3LE, UK\\
$^7$ Astrophysics Research Institute, Liverpool John Moores University, 146 Brownlow Hill, Liverpool L3 5RF, UK\\
$^8$ Instituto de Astrof\'isica de Canarias C/ V\'ia L\'actea s/n 38205 La Laguna, Tenerife, Spain\\
$^9$ Departamento de Astrof\'isica, Universidad de La Laguna, Av. del Astrof\'isico Francisco S\'anchez s/n, 38206 La Laguna, Tenerife, Spain \\
$^{10}$ Leiden Observatory, Leiden University, P.O. Box 9513, 2300 RA Leiden, the Netherlands\\
}

\date{Accepted XXX. Received YYY; in original form ZZZ}

\begin{document}

%\date{}

\label{firstpage}
\pagerange{\pageref{firstpage}--\pageref{lastpage}} 
\maketitle

\begin{abstract} 
  We examine the properties of barred disc galaxies in a $\Lambda$CDM
  cosmological hydrodynamical simulation from the EAGLE project. Our
  study follows the formation of 269 discs identified at $z=0$ in the
  stellar mass range $10.6<\log M_{*}/M_{\odot} <11$. These
  discs show a wide range of bar strengths, from unbarred discs
  ($\approx 60\%$) to weak bars ($\approx 20\%$) to strongly barred systems
  ($\approx 20\%$). Bars in these systems develop after redshift $\approx 1.3$, on
  timescales that depend sensitively on the strength of the
  pattern. Strong bars develop relatively quickly (in a few Gyr, 
  $\sim 10$ disc rotation periods) in systems that are disc
  dominated, gas poor, and have declining rotation curves. Weak bars
  develop more slowly in systems where the disc is less
  gravitationally important, and are still growing at $z=0$. Unbarred
  galaxies are comparatively gas-rich discs whose rotation speeds do
  not exceed the maximum circular velocity of the halos they
  inhabit. Bar lengths compare favourably with observations, ranging
  from 0.2 to 0.8 times the radius containing $90\%$ of the stars.
  Bars slow down remarkably quickly as they grow, causing the inner
  regions of the surrounding dark halo to expand. At $z=0$ strong bars
  have corotation radii roughly ten times the bar length. Such slow
  bars are inconsistent with the few cases where pattern speeds have
  been measured or inferred observationally, a discrepancy that, if
  confirmed, might prove a challenge for disc galaxy formation in
  $\Lambda$CDM.
 \end{abstract}

\section{Introduction}
\label{SecIntro}

%%%%%%%%%%%%%%%%%%%%%%%%%%%%%%%%%%%%%%%%%%%
\begin{figure*}
\begin{center}
\includegraphics[width=\linewidth,clip]{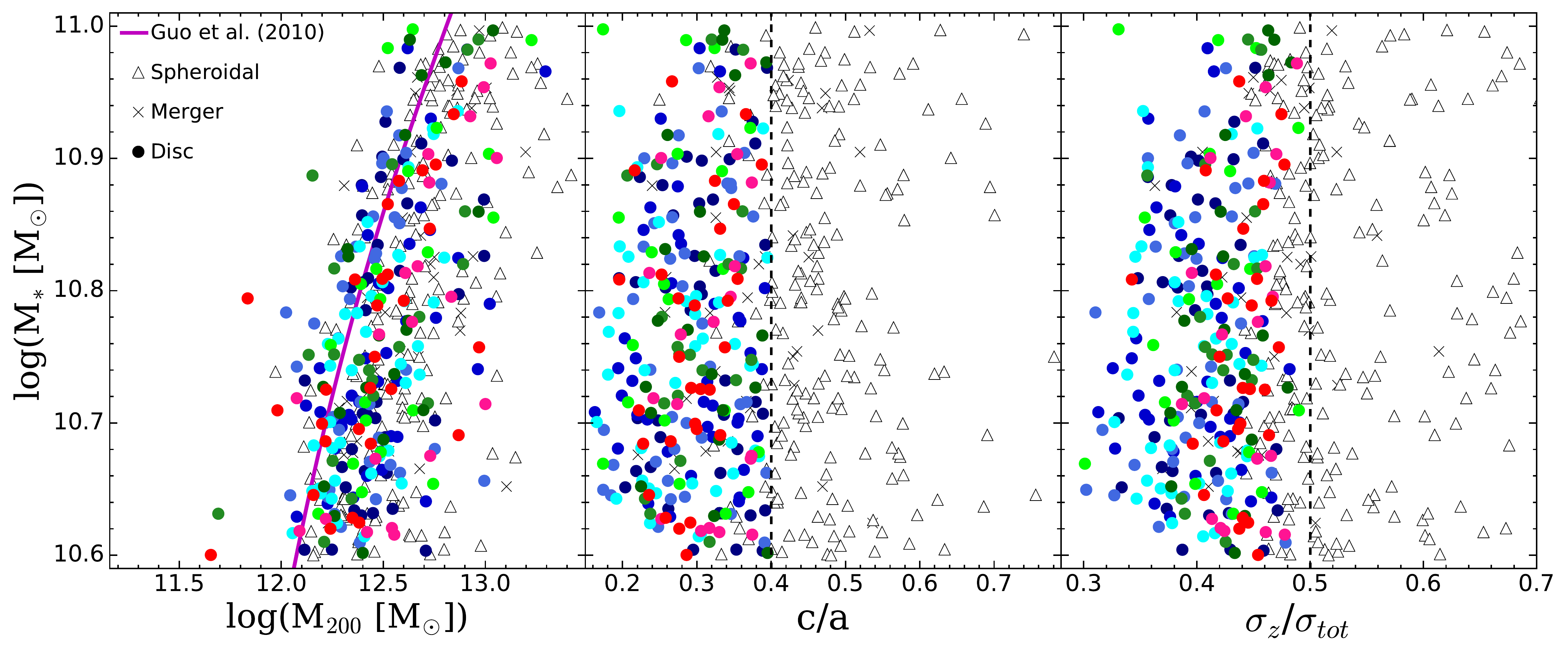}
\end{center}
\caption{Disc galaxy sample from EAGLE used in this paper. {\it Left:} Galaxy stellar
  mass, $M_{*}$, as a function virial mass $M_{200}$. Solid line
  indicates the prediction of the abundance-matching model of
  \citet{Guo2010}, for reference. {\it Middle:} Flattening parameter
  $c/a$, measured as the ratio of the eigenvalues of the principal
  axes of the inertia tensor of the stars. {\it Right:} Minor axis
  stellar velocity
  dispersion, expressed in units of the total. Vertical dashed lines indicate
  the conditions required to be selected as ``discs'' in our
  analysis. Discs are shown as coloured circles, spheroidal systems as
  open triangles, and visually identified ongoing mergers or disturbed
  systems as crosses. The colour scheme denotes the strength of the bar
  pattern (see Fig.~\ref{FigBarz0}).}
  \label{FigGxSample}
\end{figure*} 
%%%%%%%%%%%%%%%%%%%%%%%%%%%%%%%%%%%%%%%%%%%
 
The stellar discs of spiral galaxies are dynamically fragile
structures prone to morphological and dynamical transformation. These
might be triggered by external processes, such as accretion events,
mergers, or the tidal effects of satellites and neighbouring galaxies.
They may also result from internal processes, which tend to be more
subtle and to operate over longer timescales but are nonetheless
effective at inducing notable changes in the morphology and structure
of the disc. Internal processes invariably redistribute the disc's
angular momentum, driving mass inwards while pushing angular momentum
outwards.

Angular momentum redistribution requires
non-axisymmetric features \citep{LBK1972,Tremaine1984}, of which
bars---i.e., extended and radially-coherent $m=2$ perturbations to the
disc's azimuthal structure---are a particularly clear example.
Bars come in many different sizes and shapes; from short inner bars
that affect a small fraction of stars to long bars that extend out to
the confines of the disc; and from thin rectangular bars that correspond to
a single, dominant $m=2$ mode to fat oval structures with sizable
contributions from higher even Fourier modes. Taking them all
together, bars are an extremely common phenomenon in disc galaxies,
and are present in a large fraction of discs
\citep[e.g.,][]{Eskridge2000,whyte2002,marinova2007,Sheth2008,Gadotti2011}.

The origin of bars has long been an issue of debate. N-body discs
quickly turned into bars in early simulations \citep{Miller1968,
  Hockney1969}, a result that suggested a ``global instability'' that
would affect essentially all stellar discs unless stabilized by a
suitable mechanism \citep[see][for a review of early
work]{Sellwood1993}. One such mechanism was proposed by
\citet{Ostriker1973}, who argued, in an influential paper, that cold
stellar discs required the presence of a massive non-rotating dark
halo in order not to go bar unstable. In this scenario, bars 
develop quickly in systems where the disc is dominant (perhaps triggered by
accretion events or tides), whereas unbarred discs are those whose
dynamics is largely dominated by the dark halo
\citep[][]{Efstathiou1982}. This idea is still widely in use and
criteria for instantaneous ``bar instability'' are a key ingredient of
semi-analytic models of galaxy formation that attempt to match the
morphological mix of the observed galaxy population \citep[see,
e.g.,][and references therein]{Lacey2015}.

More recent work, however, has led to a more nuanced view, and it is
now recognized that bars, weak and strong, may develop gradually in most stellar
discs that are relatively massive and kinematically cold, even when
the halo is important.  Indeed, in some cases massive halos have even
been found to promote bar formation: one clear example is that
provided by \citet{Athanassoula2002a}, who shows that bars may develop
faster in disc-dominated systems, but they eventually become stronger
in halo-dominated ones. Halos apparently do not prevent bars, but,
rather, just delay their formation \citep[see][for a recent
review]{Athanassoula2013}.

Once formed, bars are a conduit for the transfer angular momentum from the
disc to other parts of the system. The more angular momentum a bar is
able to lose, the longer and thinner (``stronger'') it can become. To
grow, then, bars need material to absorb the angular momentum lost by
stars that join the bar, be it other stars in the outer disc or
particles in the halo that might get trapped in resonances with the
bar. A massive halo can therefore aid this process by providing a sink
for the angular momentum lost by stars that make up the growing bar
\citep{Athanassoula2003}.

Bars can therefore develop gradually over many orbital periods, on a
timescale that depends mainly on the relative importance of disc vs
halo, but likely influenced as well by the velocity dispersion of the
disc (hotter discs are less prone to global distortions) and by the
potential well depth of the halo (faster moving halo particles are
harder to trap into resonances with the bar). The two
scenarios---instantaneous bar instability vs gradual bar
growth---should in principle yield different predictions for the
abundance, size, and pattern speeds of bars, as well as for their
evolution with redshift, but detailed predictions in a proper
cosmological setting have yet to be worked out.

One corollary of gradual bar formation is that bars that grow
longer/stronger should slow down
\citep{Hernquist1992,Debattista2000}. This is because bars cannot
extend beyond corotation, the radius where the angular speed of a
circular orbit equals that of the bar pattern
\citep{Contopoulos1980}. Angular speeds decrease outwards, so the
longer the bar grows the slower its pattern speed must become. The bar
cannot grow longer than the disc, of course, but it can continue to
slow down, implying that the ratio between corotation radius
($r_{\rm corot}$) and bar length ($l_{\rm bar}$) can provide interesting
constraints on the relative importance of the disc and halo, as well
as on the time elapsed since the onset of the bar
\citep{Debattista2000}. Although the measurements are challenging and
often indirect, most observational estimates point to ``fast bars''
where $r_{\rm corot}<1.4\, l_{\rm bar}$ \citep[see,
e.g.,][]{Elmegreen1996,Corsini2011}.

Interestingly, bar slowdown might also have discernible effects on the
dark matter density profile, since it is the halo that absorbs much of
the disc angular momentum, especially in the case of strong bars. A
number of studies have indeed suggested that the central density
cusps expected in cold dark matter halos \citep{Navarro1996,Navarro1997}
might be softened and perhaps erased\footnote{Others, however, have
  argued otherwise, so the issue is still under debate
  \citep{Sellwood2003,Sellwood2008,Dubinski2009}.} by a bar
\citep{Weinberg2002,Holley-Bockelmann2005}. This result
has important consequences for models of gamma ray emission by dark
matter annihilation in the direction of the Galactic centre
\citep{Schaller2016}: the Milky Way is, after all, a barred galaxy
\citep[e.g.,][]{Blitz1991}.

The discussion above suggests that the abundance of barred galaxies,
together with the distribution of bar strengths, lengths, and pattern
speeds, may provide interesting constraints on the mass, size, and
kinematics of disc galaxies, on the time of their assembly, and on the
mass and density profiles of the dark matter halos they inhabit.  This
is important, because, once a cosmological model has been adopted, the very
same properties that govern bar growth are independently specified by
other constraints, and cannot be tuned arbitrarily. Success in
reproducing the properties of the barred galaxy population in a
particular cosmology is thus far from assured.

 %%%%%%%%%%%%%%%%%%%%%%%%%%%%%%%%%%%%%
\begin{figure*}
\begin{center}
\includegraphics[width=\linewidth,clip]{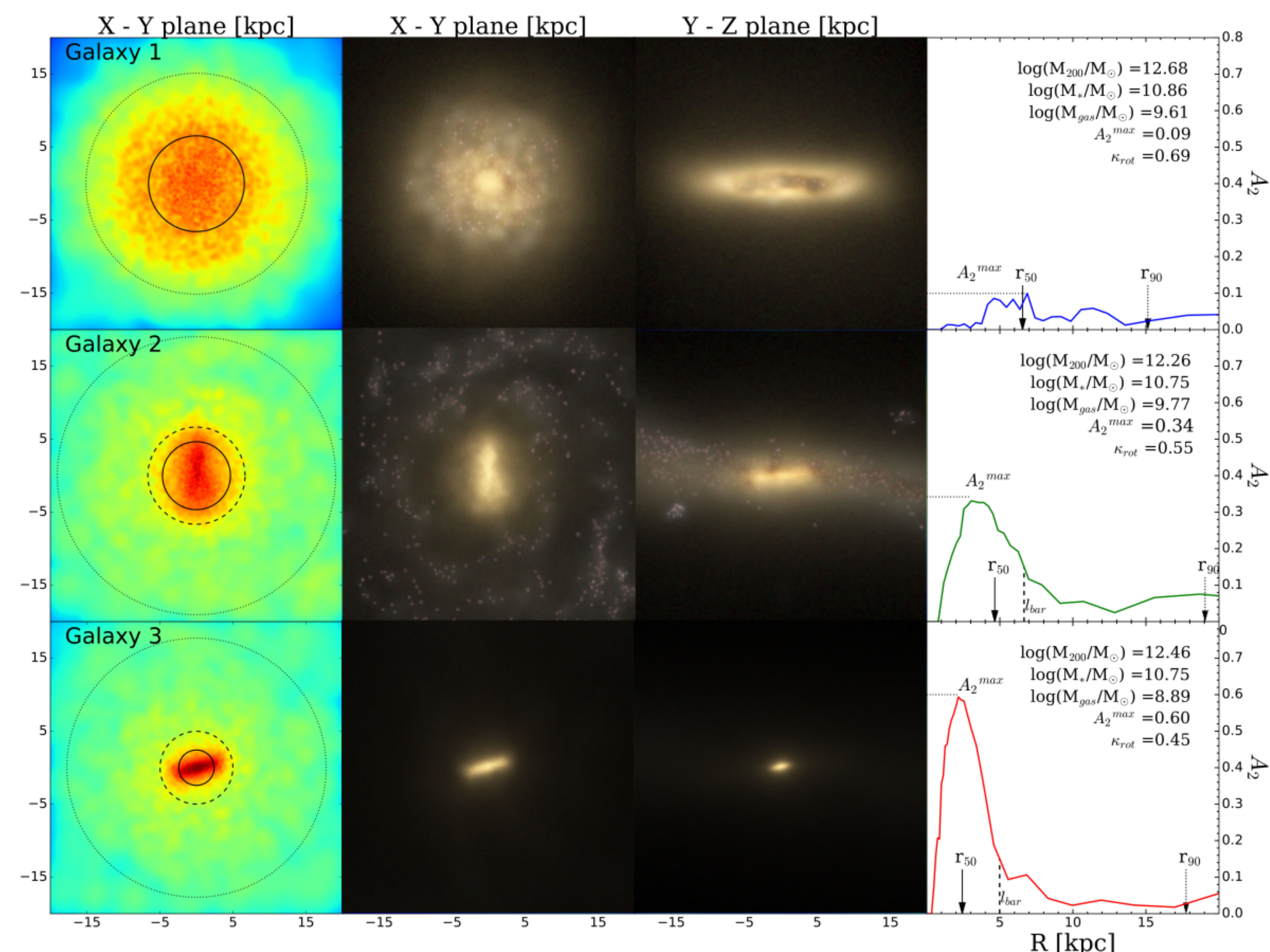}
\end{center}
\caption{Projected stellar density maps for three examples of an
  unbarred galaxy (top row), a weak bar (middle row) and a strongly barred
  disc (bottom row). The leftmost column shows face-on views of the
  three galaxies. %created with the code py-sphviewer \citep{Benitez-Llambay2015}.
   Dotted, dashed, and solid circles on the images indicate the
  galaxy radius, $r_{\rm 90}$, the bar length, $l_{\rm bar}$, and the
  stellar half-mass radius, $r_{50}$.
    The middle and right panels show face-on and edge-on views, respectively,
  created with the radiative transfer code SKIRT \citep{Baes2011}.
  These images show the stellar light based on monochromatic SDSS
$u$, $g$ and $r$ band filters and accounting for dust extinction.
The rightmost column shows the radial profile of the
  bar strength parameter, $A_2(R)$, and indicates a few characteristic
  radii. 
}
\label{FigGxImages}
\end{figure*} 
%%%%%%%%%%%%%%%%%%%%%%%%%%%%%%%%%%%%%%%

In $\Lambda$CDM models---the current paradigm of
structure formation---the relation between galaxy mass and halo mass
may be derived using ``abundance
matching'' arguments \citep{Frenk1988,Vale2006,Guo2010,Moster2013,Behroozi2013}. Further,
galaxy sizes are also constrained by scaling laws such as the
Tully-Fisher relation \citep[see, e.g.,][]{Ferrero2016}. Do galaxies that match
those constraints also result in a barred galaxy population whose
statistics, bar lengths, and pattern speeds are compatible with
observation? 

We address this question here by examining the properties of disc
galaxies in the EAGLE cosmological hydrodynamical simulation of a
$\Lambda$CDM universe \citep{Schaye2015,Crain2015}. This paper is
organized as follow. In Sec. \ref{SecNumSim} we briefly describe the
numerical simulations and the galaxy sample
selection. Sec.~\ref{SecRes} presents the results of our analysis,
including the frequency of bars (Sec.~\ref{SecBarFreq}); bar lengths
(Sec.~\ref{SecBarL}); bar growth (Sec.~\ref{SecBarG}); bar slowdown
(Sec.~\ref{SecBarS}); and its effects on the halo mass profile
(Sec.~\ref{SecHaloEvol}).  We summarize our main conclusions in
Sec.~\ref{SecConc}.

%%%%%%%%%%%%%%%%%%%%%%%%%%%%%%%%%%%%%
\begin{figure}
\begin{center}
\includegraphics[width=\linewidth,clip]{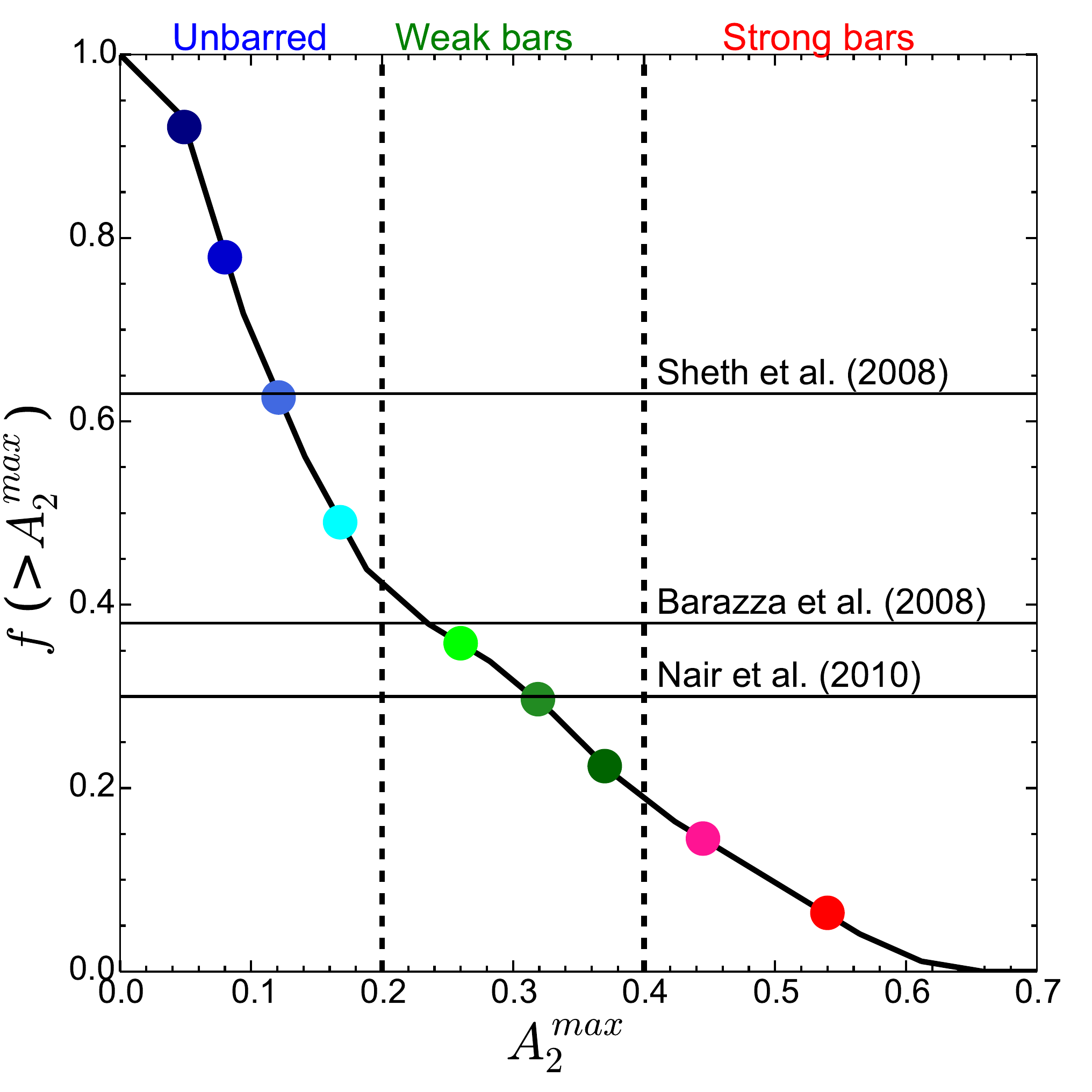}
\end{center}
\caption{Cumulative distribution of the bar strength parameter
  $A_2^{\rm max}$, compared with observational estimates of the bar
  fraction of galaxies with comparable stellar mass.
  The colour scheme assigns different hues of red to
  strong bars ($A_{2}^{\rm max}>0.4$), of green to weak bars
  ($0.2<A_{2}^{\rm max}<0.4$), and of blue to unbarred systems ($A_{2}^{\rm max}<0.2$).}
   \label{FigBarFreq}
\end{figure} 
%%%%%%%%%%%%%%%%%%%%%%%%%%%%%%%%%%%%%%%

\section{Numerical Simulation and Sample Selection}
\label{SecNumSim}

\subsection{The EAGLE simulation}
\label{SecEAGLE}

We use galaxies identified in one of the cosmological hydrodynamical
simulations of the EAGLE (Evolution and Assembly of GaLaxies and their
Environments) Project. This run, labelled ``Ref-L100N1504'' in
\citet{Schaye2015}, follows the evolution of $2 \times 1504^3$
particles (baryons $+$ dark matter) in a large cosmological box with $100$ comoving Mpc on a
side, adopting a flat $\Lambda$CDM cosmology consistent with the
cosmological parameters from the \citet{PlankCollaboration2014}:
$H_{0}$ = 67.77 km s$^{-1}$, $\sigma_{8} $= 0.8288, $n_{\rm s}$=0.9611,
$\Omega_{m}$=0.307, $\Omega_{\Lambda}$=0.693 and
$\Omega_{\rm b}$=0.04825. 

Initial conditions were generated using second
order Lagrangian perturbation theory \citep{Jenkins2010} for gas and
dark matter particles with masses equal to
$1.81 \times 10^6$ M$_{\odot}$ and $9.70 \times 10^6$ M$_{\odot}$,
respectively.  The Plummer-equivalent gravitational softening is
set to $2.66$ comoving kpc and capped to a maximum physical value of 0.7 kpc at $z=2.8$.  Full
particle properties are recorded at 29 snapshots between redshift 20
and 0. In addition, a reduced set of particles properties are recorded
at 405 outputs between redshift 20 and 0. For a detailed description
of the simulations we refer the reader to \citet{Schaye2015}.

The simulation was performed using a modified version of the
{\sc gadget3} code, a descendant of {\sc gadget2}
\citep{Springel2005b}, with a version of Smoothed Particle
Hydrodynamics (SPH) technique %\citep[]{Gingold1977, Lucy1977} 
modified to a pressure-entropy formulation of the equations of motion
\citep{Hopkins2013,Schaller2015}.  The simulation includes a prescription for
radiative cooling and heating implemented following
\citet{Wiersma2009a}.

Star formation is treated stochastically following the
pressure-dependent Kennicutt-Schmidt relation
\citep{Schaye2008}, with a metal-dependent density
threshold \citep{Schaye2004}. The stellar initial mass function is
assumed to be that of \citet{Chabrier2003} and the stellar mass loss
is modelled \citep{Wiersma2009}. Feedback from star formation is
implemented thermally and stochastically following
\citet{DallaVecchia2012}. Black holes growth is modelled using a
modified version of the Bondi-Hoyle accretion and can input energy to
their surrounding gas through AGN feedback \citep{RosasGuevara2012}.

The subgrid
parameters of EAGLE have been calibrated to match the galaxy stellar
mass function and the average size of galaxies as a function of mass
at $z=1$. Earlier papers have demonstrated that EAGLE broadly
reproduces a number of well known properties of the galaxy population,
including their colours, metallicities, alpha-enhancement, star
formation rates, gas content, and scaling laws
\citep{Furlong2015,Lagos2015,Rahmati2015,Schaller2015a,Schaye2015,
  Trayford2015,Trayford2016,Bahe2016,Camps2016,Segers2016}.

\subsection{Galaxy sample}
\label{SecGxSample}

Dark matter haloes are identified at every snapshot using a
friends-of-friends (FoF) algorithm  with linking
length equal to 0.2 times the mean interparticle separation \citep{Davis1985}.
Baryonic particles are then assigned to the same FoF halo as their closest dark
matter neighbour. Gravitationally bound subhaloes are then identified
in each FoF halo using the {\sc SUBFIND} algorithm
\citep{Springel2001a,Dolag2009}. We shall only consider the most
massive (``central'') subhalo of each FoF grouping and neglect
``satellite'' galaxies in the analysis that follows.

Our sample includes systems within a narrow range of stellar mass,
$10.6 \le \rm log(M_{*}/M_{\odot}) \le 11$, measured within a
sphere of $30$ kpc radius centred at the potential minimum of the
halo. We shall refer to the radius containing half of all stars as
$r_{50}$ and that containing $90\%$ of all stars as $r_{90}$. 

We focus our analysis on individual galaxies resolved with at least
20,000 star particles and more than 100,000 dark matter particles, so
as to be able to discern their morphological traits (discs, spheroids,
bars) and measure their internal structure. This resolution is at the
limit of what is currently achievable in simulations that aim to
resolve the galaxy population in a cosmologically significant
volume. Although it is by no means ideal to follow in detail the
intricate internal dynamics governing the evolution of barred galaxies
\citep[many authors argue that many millions of particles per galaxy
are required, see, e.g.,][]{Weinberg2007a}, we believe that this is
still an instructive exercise, especially because ours is one of the
first studies of barred galaxies {\it as a population} in a proper
cosmological setting. Earlier work has mainly focussed on ``zoom-in''
simulations of individual systems
\citep{Curir2006,Scannapieco2012,Kraljic2012,Guedes2013,Okamoto2014,Goz2015}
and can therefore not address the questions we pose here.

%%%%%%%%%%%%%%%%%%%%%%%%%%%%%%%%%%%%%%%%%%%
\begin{figure*}
\begin{center}
\includegraphics[width=\linewidth,clip]{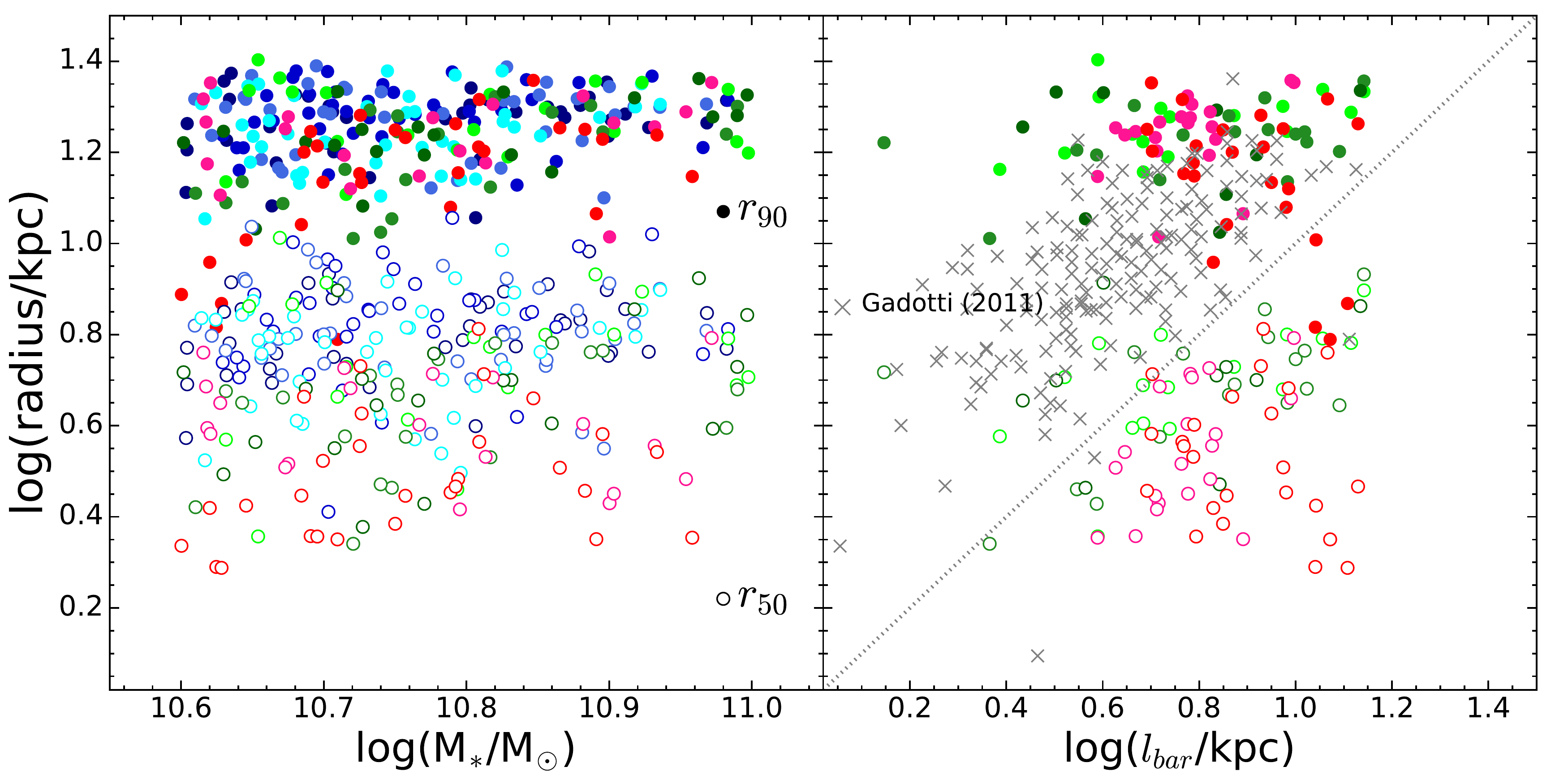}
\end{center}
\caption{Bar lengths and disc characteristic radii. {\it Left:}
  Stellar half-mass radius and $90\%$-mass radius as a function of
  galaxy stellar mass, $M_{*}$. {\it Right:} Bar length compared
with disc radii for galaxies with $A_{2}^{\rm max}>0.2$. 
Colours indicate bar strength, as in
Fig.~\ref{FigBarFreq}. Grey symbols indicate $l_{\rm bar}$ vs $r_{90}$ results from
the SDSS observations of \citet{Gadotti2011}, for reference. Note that most of his galaxies are
smaller (less massive) than those in our sample. }
\label{FigBarL}
\end{figure*} 
%%%%%%%%%%%%%%%%%%%%%%%%%%%%%%%%%%%%%%%%%%%

There are 495 central galaxies that satisfy our selection criteria at
$z=0$. Fig. \ref{FigGxSample} shows some of the properties of
this sample. As a function of stellar mass, the left panel shows the
virial\footnote{Throughout this paper, virial quantities are computed
  within radius where the enclosed density is $200$ times the critical
  density of the Universe and are denoted by a ``200'' subscript.}
mass of the system; the middle panel shows its flattening, using the
axis ratio $c/a$ of the inertia tensor principal axes; whereas the
right-hand panel shows the vertical (i.e., along the axis parallel
to the stars' angular momentum) velocity dispersion, $\sigma_z$, in
units of the total velocity dispersion of stars in the system,
$\sigma_{\rm tot}$.

As shown by the solid line in the left-hand panel of
Fig.~\ref{FigGxSample} {\sc eagle} central galaxies in this mass range
follow roughly the abundance-matching relation expected between
$M_{*}$ and $M_{200}$ from the model of \citet{Guo2010}. We
identify discs as flattened systems kinematically cold in the vertical
direction that satisfy simultaneously the following two conditions:
$c/a<0.4$ and $\sigma_z<0.5\, \sigma_{\rm tot}$. Discs are shown as
coloured filled circles in Fig.~\ref{FigGxSample}; other galaxies are
shown as either open triangles (``spheroidals'') or crosses for
ongoing mergers identified through individual visual inspection. Our
final sample contains 269 discs, 193 spheroidals, and 33 ongoing
mergers. We shall only consider discs in the analysis that follows.

We divide the sample of discs into three categories: different hues of
red for strong bars, of green for weak bars, and of blue for unbarred
systems (Sec.~\ref{SecBarID}). Note that in the right-hand panel of Fig.~\ref{FigGxSample}
strong bars tend to be located in discs with higher vertical velocity
dispersion.  This is likely the result of the three-dimensional
structure of the bars and of the effect of the ``buckling
instability'' \citep{Raha1991}, which tend to make discs thicker
after a bar forms.

%%%%%%%%%%%%%%%%%%%%%%%%%%%%%%%%%%%%%%%%%%%%%%%%%%%%%%%%%%%%%%%
\begin{figure}
\begin{center}
\includegraphics[width=0.9\linewidth,clip]{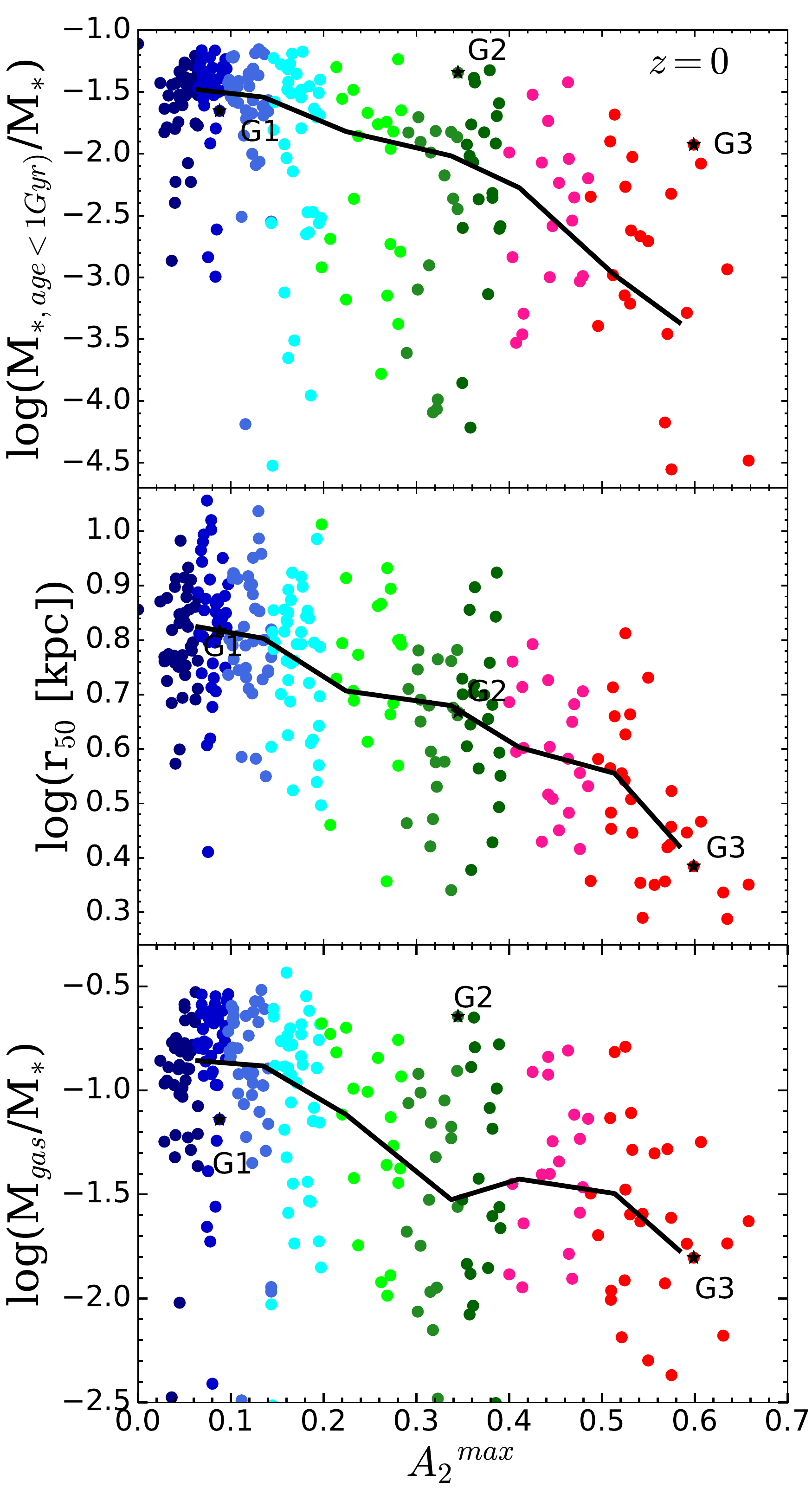}
\end{center}
\caption{Bar strength parameter $A_{2}^{\rm max}$ as a function of gas
  mass fraction (bottom), of half-mass radius (middle) and of the stellar mass form in the last
  Gyr at $z=0$ for all discs in our sample. Solid lines trace the median as a function
  of bar strength. G1, G2 and G3 refer to Galaxies 1, 2 and 3 in Fig.~\ref{FigGxImages}, respectively. }
   \label{FigBarz0}
\end{figure} 
%%%%%%%%%%%%%%%%%%%%%%%%%%%%%%%%%%%%%%%%%%%%%%%%%%%%%%%%%%%%%%%

\section{Results}
\label{SecRes}

\subsection{Bars in simulated discs}
\label{SecBarID}

We identify bars in our simulated discs by measuring the amplitude of
the $m=0$ and $m=2$ Fourier modes of the azimuthal distribution of
disc particles in the plane perpendicular to the angular momentum
vector of stars in the galaxy.  In practice, we measure
\begin{equation}
 a_{m} (R) = \sum^{N_{R}}_{i=1} M_{i} \, {cos}(m \, \phi_i ),
 \end{equation}
and
\begin{equation}
 b_{m} (R)= \sum^{N_{R}}_{i=1} M_{i} \, {sin}(m \, \phi_i ),
\end{equation}
where $N_{R}$ is the number of stellar particles in a given
cylindrical annulus of mean radius $R$, $M_{i}$ is the mass of the
$i$-th particle and $\phi_{i}$ is its azimuthal angle
\citep{Athanassoula2012}.

We use the ratio
\begin{equation}
 A_2(R) = \frac{\sqrt{a^2_2 + b^2_2}}{a_{0}},
\end{equation}
to measure the strength of the $m=2$ mode, and shall use its maximum
value $A^{\rm max}_{2}=$ max($A_2(R)$) as a measure of the strength of
the bar component.

%%%%%%%%%%%%%%%%%%%%%%%%%%%%%%%%%%%%%
\begin{figure}
\begin{center}
\includegraphics[width=\linewidth,clip]{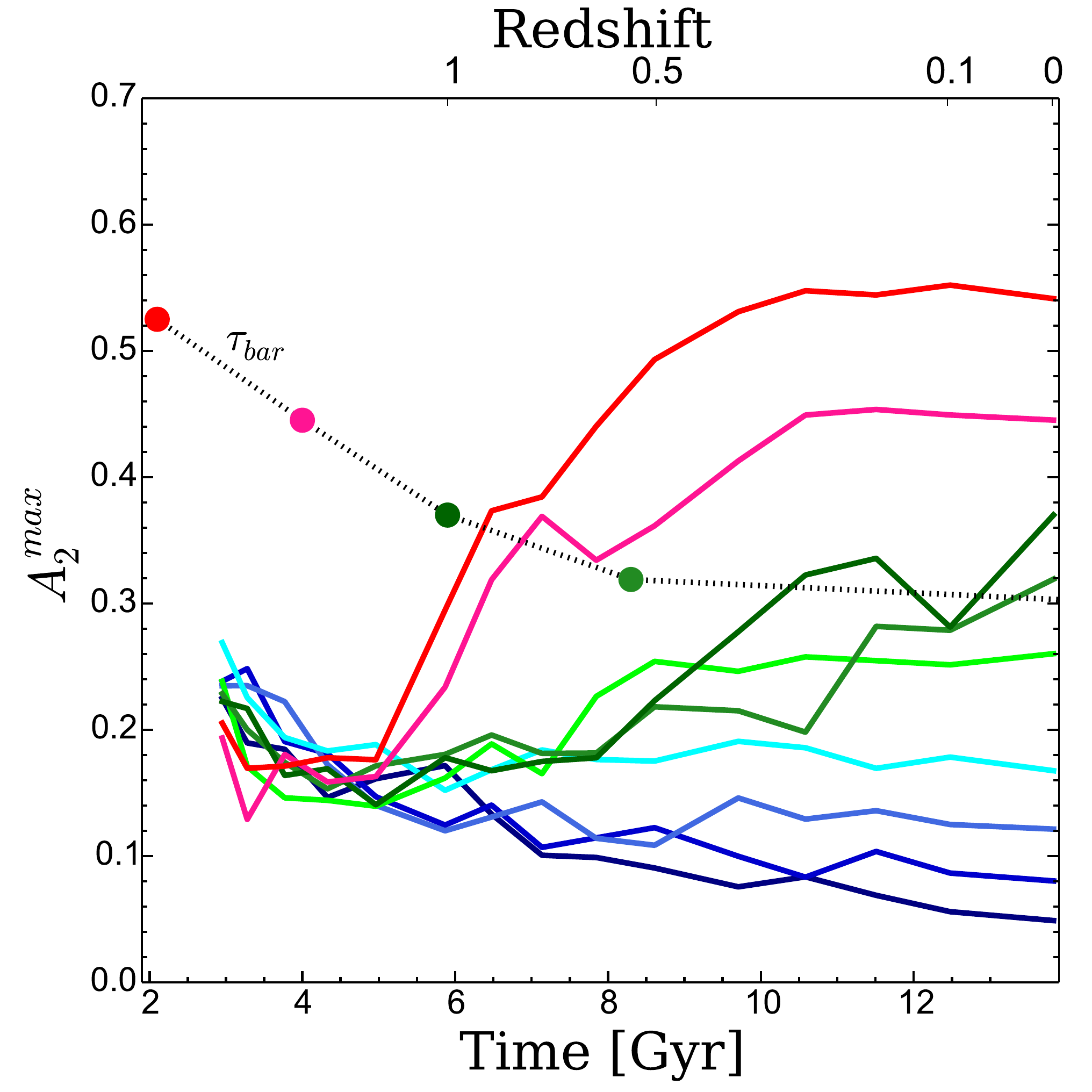}
\end{center}
\caption{Evolution of the bar strength parameter, averaged in bins of
  galaxies according to their value of $A_2^{\rm max}$ at
  $z=0$. Colour scheme for the curves is as in Fig.~\ref{FigBarFreq}. Bar
growth timescales (defined arbitrarily as the time it takes to increase the bar
strength from $0.2$ to $0.4$) are shown by the grey connected
symbols. Note that this approaches the Hubble time for the weakest bars.}
 \label{FigBarG}
\end{figure} 
%%%%%%%%%%%%%%%%%%%%%%%%%%%%%%%%%%%%%%%

Fig. \ref{FigGxImages} shows projected stellar density maps for three
simulated galaxies with different values of $A^{\rm max}_2$. The top
row shows a galaxy with $A^{\rm max}_2\approx 0.1$, where no obvious
bar is present (Galaxy 1). The middle row shows a system with a clear
oval structure resembling a weak bar (Galaxy 2; $A^{\rm max}_2\approx
0.35$). Finally, the bottom row shows a strongly barred case, where
$A^{\rm max}_2\approx 0.6$ (Galaxy 3). We shall hereafter use the
value of $A^{\rm max}_2$ to classify galaxies as unbarred ($A^{\rm
  max}_2 <0.2$) , weakly barred ($0.2<A^{\rm max}_2 <0.4$), and
strongly barred ($A^{\rm max}_2 >0.4$). Although $A^{\rm max}_{2}$
could in principle also be large for two-armed spirals, these
typically peak at lower values than those we have used to define
bars. We have visually checked every galaxy to make sure that our
barred galaxies do not include spurious cases.

The right-hand column of Fig. \ref{FigGxImages} shows the radial
profile of $A_2$ for the three examples, and indicates a few
characteristic radii: $r_{50}$, $r_{90}$, and the bar length, $l_{\rm
  bar}$, which we define as the radius where the $A_2$ profile first
dips below $0.15$ after reaching its peak. Various circles indicate
these radii on the galaxy images; note that this definition of $l_{\rm
  bar}$ (dashed circles) coincides well with the radial extent of the
bar, as measured from the face-on map of the stellar distribution.

Finally, we shall use $\phi_{\rm bar}=0.5\tan^{-1}(b_2/a_2)$, measured at the
radius where $A_2(R)$ peaks, to define the bar position angle. The
time variation of this angle is used to estimate the bar
pattern speed in the analysis that follows.

\subsection{Bar frequency}
\label{SecBarFreq}

Although qualitatively there is broad consensus that bars are
relatively common, quantitatively there is less agreement on the
fraction of discs that are barred. This is a result of several
factors, including the facts that (i) there is no standard definition
of what constitutes a bar; that (ii) bar prominence depends on
wavelength \citep[stronger in the infrared; e.g.,][]{Eskridge2000},
morphological type \citep[longer in early-type spirals;
e.g.,][]{Elmegreen1985}, galaxy mass \citep[decreasing with increasing
mass; e.g.,][]{Nair2010}, and redshift \citep[less frequent at early
times; e.g.,][]{Sheth2008}; and that (iii) various studies differ on how to
define the parent population of discs (would a galaxy be classified as
barred or as a spheroid if it had no obvious disc component?).

Although these shortcomings hinder a definitive comparison of
our results with observations, we contrast our findings
with a few recent observational estimates in
Fig.~\ref{FigBarFreq}. This figure shows the cumulative distribution
of our bar strength parameter $A_2^{\rm max}$ and compares it with
the bar fraction estimates of three studies that report bar fractions
as a function of galaxy mass.

More specifically, \citet{Barazza2008} report a fraction of $\approx
38\%$ for galaxies in the mass range considered in our
analysis. On the other hand, \citet{Sheth2008} find, at low
redshift and in the same mass range, a much higher bar fraction of
$\approx 62\%$. Finally,  \citet{Nair2010} report a much lower bar
fraction of only about $30\%$ in a comparable mass range. Given these
disparate estimates, our finding that about $40\%$ of EAGLE discs have
bars (weak or strong) seems quite consistent with observations.

\subsection{Bar lengths}
\label{SecBarL}

As discussed in Sec.~\ref{SecIntro}, the length of a bar is an
important parameter characterizing the evolutionary stage of the bar
phenomenon. Bars are expected to grow longer with time so, if bars
were triggered too early and/or their growth timescales were too
short, then bar lengths, when expressed in units of the disc
characteristic radii, would be too long. Such concerns have been
raised by \citet{Erwin2005}, for example, who argue that there is a
shortage of ``short bars'' in simulations.

We examine the length of bars in Fig.~\ref{FigBarL},
where the left panel shows, as a function of the galaxy stellar mass,
the half-mass radius and $90\%$-mass radius of all EAGLE discs. The
right-hand panel shows the same radii, but as a function of bar
length, for all barred EAGLE galaxies ($A_2^{\rm max}>0.2$). Note that bar lengths tend to
straddle both radii; in general $r_{50}<l_{\rm bar}<r_{90}$. Relative
to $r_{90}$, bar lengths span the whole available range: the shortest
bars have $l_{\rm bar} \sim 0.15\, r_{90}$; the longest reach
$l_{\rm bar}=r_{90}$ and even exceed it in a few cases. There is
certainly not a shortage of short bars in our simulations, at least as
measured by $l_{\rm bar}/r_{90}$.

The EAGLE predictions are in broad agreement with the results of \citet{Gadotti2011},
who show that the distribution of bar lengths in his sample of nearly
1,000 SDSS galaxies peaks at roughly $\sim 0.6\, r_{90}$. In addition,
there are few bars shorter than $0.25\, r_{90}$ or longer than
$0.85\, r_{90}$. His results for $r_{90}$ vs $l_{\rm bar}$ are shown
in the right-hand panel of Fig.~\ref{FigBarL} with grey
crosses. Although many galaxies in the observed sample are less
massive, and thus smaller than ours, it is clear that there is no
obvious discrepancy between observations and simulations in the regime
where they overlap in $r_{90}$. We conclude that bar lengths in EAGLE
galaxies are well within the range allowed by observations.

\subsection{Barred galaxy properties at $z = 0$}
\label{SecBarProp}

We now explore the relation (at $z=0$) between bar strength and the properties
 of the discs in which they form. Fig.~\ref{FigBarz0} shows
that bars are stronger in discs that are more centrally concentrated
(i.e., smaller half-mass radii), that bars are relatively gas poor,
and that they have formed fewer stars in the past Gyr than other
discs.

Indeed, unbarred galaxies in our sample are typically forming stars at
rates roughly about $40\%$ of their past average. However, star
formation rates decrease strongly with increasing bar strength, to
roughly $1\%$ of the past average for the strongest bars.  A related
result is that barred discs differ strongly from unbarred ones in
their star formation history. Indeed, on average, 50\% (90\%) of all
stars in our strong bars have formed by cosmic time $t=3.76$ ($6.15$) Gyr, compared
with $t=5.45$ ($11.06$) Gyr for unbarred systems.

As shown by Fig.~\ref{FigBarz0}, strongly-barred discs are roughly
three times smaller than unbarred systems of similar stellar
mass. This is an important clue that bar formation proceeds more
rapidly in systems where the stellar component is more gravitationally
dominant. We turn our attention to the timescale of bar growth next.

\subsection{Bar growth}
\label{SecBarG}

Fig.~\ref{FigBarG} shows the evolution of the bar strength parameter,
averaged for galaxies binned as a function of their value of
$A_2^{\rm max}$ at $z=0$, for the sake of clarity. This shows that bars
have developed in these systems only in the past 8 Gyrs; indeed, at
$z\sim 1.3$ ($t\sim 5$ Gyr) very few, if any, of the present day EAGLE
discs in our sample had a measurable bar. Note that this statement
only applies to the current sample, and should {\it not} be understood
as implying that the bar fraction in EAGLE necessarily declines with
redshift. We intend to address that topic in future work, but restrict
ourselves here to the evolution of $z=0$ EAGLE discs within a narrow
range of stellar mass, as described in Sec.~\ref{SecGxSample}.

Fig.~\ref{FigBarG} illustrates a number of interesting points: (i)
bars, especially strong ones, are in general not a recurrent
phenomenon; (ii) strong bars develop quickly and saturate, whereas
weak bars are still growing at $z=0$; (iii) few unbarred galaxies have
had bars in the past; and, finally, (iv) the timescale for bar growth
is clearly a strong function of final bar strength. (We have
explicitly checked that none of these conclusions are a result of the
averaging procedure.) We illustrate this by the dotted line in
Fig.~\ref{FigBarG}, which indicates the timescale $\tau_{\rm bar}$
(defined as the time needed for a bar to grow from $A_2^{\rm max}=0.2$
to $0.4$) as a function of final bar strength.

Note that even strong bars grow over several Gyr, or tens of half-mass
disc rotation periods\footnote{The average disc rotation period of
  discs in our sample at $r=5$ kpc is $0.14$ Gyr.}. Weak bars take much
longer to develop. We conclude that bars in EAGLE discs are best
described as developing gradually over many rotation periods rather
than as the result of a ''global instability'' that proceeds nearly
instantaneously when triggered.

What sets the bar growth timescale? Or, more generally, what parameter
best predicts the development of a bar? A clue may be gleaned from
Fig.~\ref{FigBarz0}, where we showed that barred discs are on average
more centrally concentrated that unbarred ones: this is in agreement
with the findings of earlier work which suggested that
gravitationally-dominant discs are the ones where bars will grow
faster (see Sec.~\ref{SecIntro}). A simple quantitative estimate is
given by the ratio between the circular velocity at the half-mass
radius, $V_{50}=V_c(r_{50})$, and the disc contribution,
$V_{\rm disc}=(GM_{*}/r_{50})^{1/2}$,
\begin{equation}
f_{\rm disc} \equiv \frac{V_{50}}{V_{\rm disc}}.
\label{EqFdisc}
\end{equation}
This type of formulation was first proposed by \citet{Efstathiou1982}
and is the one usually adopted in semi-analytic models such as GALFORM
\citep[see, e.g.,][]{Cole2000,Bower2006,Lacey2015}. These models
typically assume that bars develop in discs with $f_{\rm disc}<1.1$,
and that others remain unbarred.

The $f_{\rm disc}$ parameter measures the {\it local} importance of
the disc but there is evidence to suggest that, on its own, it is
insufficient to predict which galaxies will become barred. As
discussed by \citet{Athanassoula2002} and \citet{Athanassoula2003},
simulations show not only that some ``$f_{\rm disc}$-stable'' discs
may become barred, but also that presumably unstable,
low-$f_{\rm disc}$ systems may be stabilized when placed within
massive halos of high velocity dispersion.  In other words, what
matters is not just the {\it local} gravitational importance of the
disc, but also its global importance to the whole system, including
its halo.

%%%%%%%%%%%%%%%%%%%%%%%%%%%%%%%%%%%%%
\begin{figure}
\begin{center}
\includegraphics[width=\linewidth,clip]{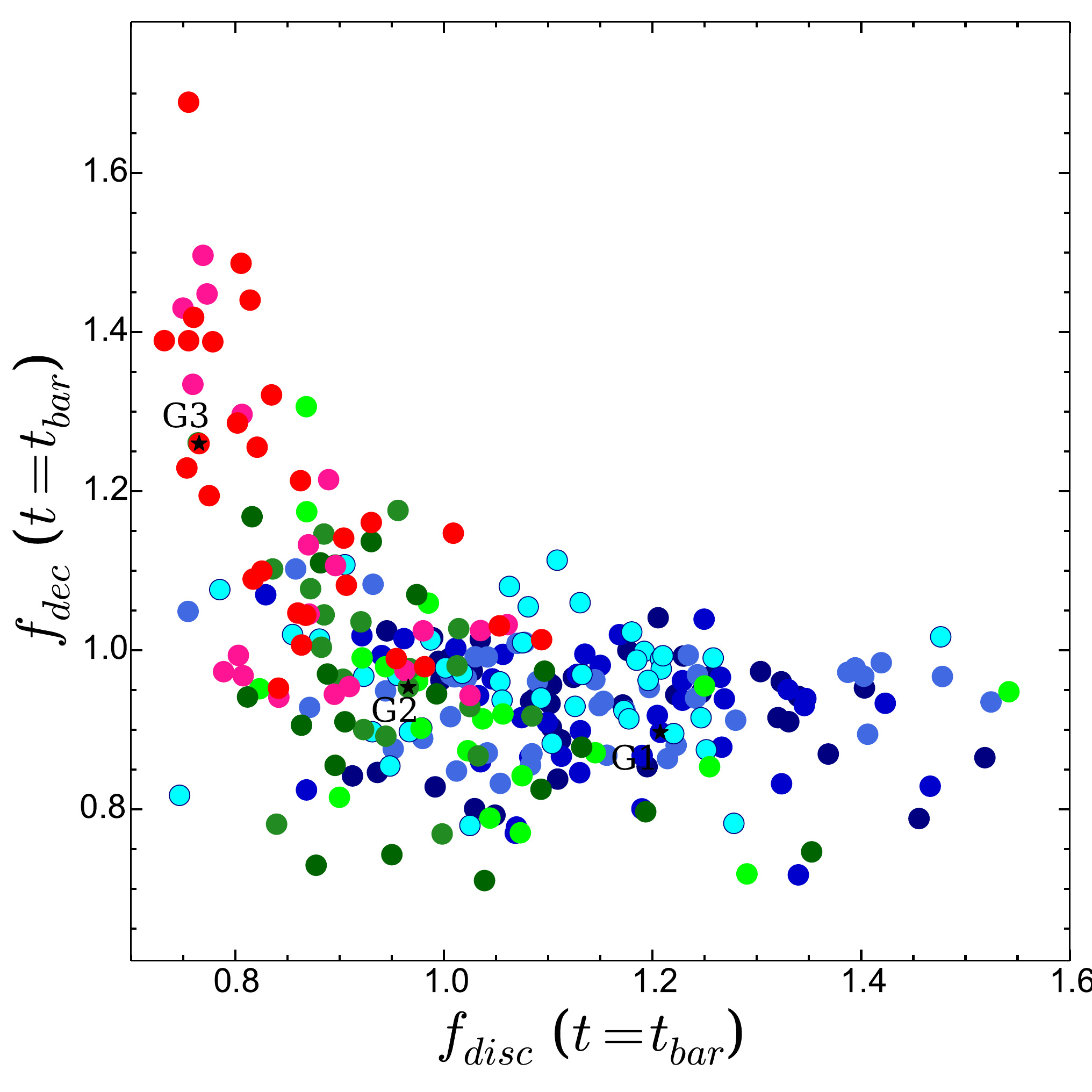}
\end{center}
\caption{Disc gravitational importance at $t=t_{\rm bar}$, defined as
  the time just before bars form (for weak and strong bars) or at
  $z=0.5$ for unbarred discs. The parameter $f_{\rm disc}$ measures
  the contribution of the disc to the circular velocity at the
  half-mass radius (Eq.~\ref{EqFdisc}). The parameter $f_{\rm dec}$ measures the global
  importance of the disc to the system (Eq.~\ref{EqFdec}). Colour scheme is as in
  Fig.~\ref{FigBarFreq}. Combining $f_{\rm dec}$ with $f_{\rm disc}$
  improves predictions of which discs will develop bars. See the text for
  definitions and further discussion. }
   \label{FigFdiscFdec}
\end{figure} 
%%%%%%%%%%%%%%%%%%%%%%%%%%%%%%%%%%%%%%%

A crude measure of the latter is provided
by the ratio between the circular velocity at the half mass radius, 
$V_{50}$, and the maximum circular velocity of the surrounding halo 
(which typically peaks far outside the disc),
\begin{equation}
f_{\rm dec} = \frac {V_{50}}{V_{\rm max,halo}}.
\label{EqFdec}
\end{equation}

With this definition, systems with $f_{\rm dec}<1$ are those whose
circular velocity curves rise beyond the outer confines of the
disc. The smaller $f_{\rm dec}$ the higher the velocities of halo
particles are relative to the disc, which may prevent them from
coupling effectively to the bar, delaying its onset or averting it
altogether. Systems with $f_{\rm dec}>1$, on the other hand, are those
with ``declining'' rotation curves, where the disc is dominant and its
rotation speed is high compared with the speed of most halo particles.

We examine this in Fig.~\ref{FigFdiscFdec}, where we show $f_{\rm
  disc}$ vs $f_{\rm dec}$ for all galaxies in our sample, measured
just {\it before}\footnote{In practice, we choose $t_{\rm bar}$ as the
  time when $A_2^{\rm max}$ first exceeds $0.2$. We set $t_{\rm
    bar}=8.6$ Gyr ($z=0.5$) for unbarred systems.} the bar develops,
at $t=t_{\rm bar}$. This figure shows clearly that the $f_{\rm
  disc}<1.1$ criterion does not accurately predict which galaxies will
become barred: $45\%$ of discs satisfying this criterion remain
unbarred, and most such discs ($73\%$, to be more precise) have
``rising'' circular velocity curves, i.e., $f_{\rm dec}<1$.

%%%%%%%%%%%%%%%%%%%%%%%%%%%%%%%%%%%%%%%%%%%%%%%%%%%%%%%%%%%%%%%
\begin{figure}
\begin{center}
\includegraphics[width=\linewidth,clip]{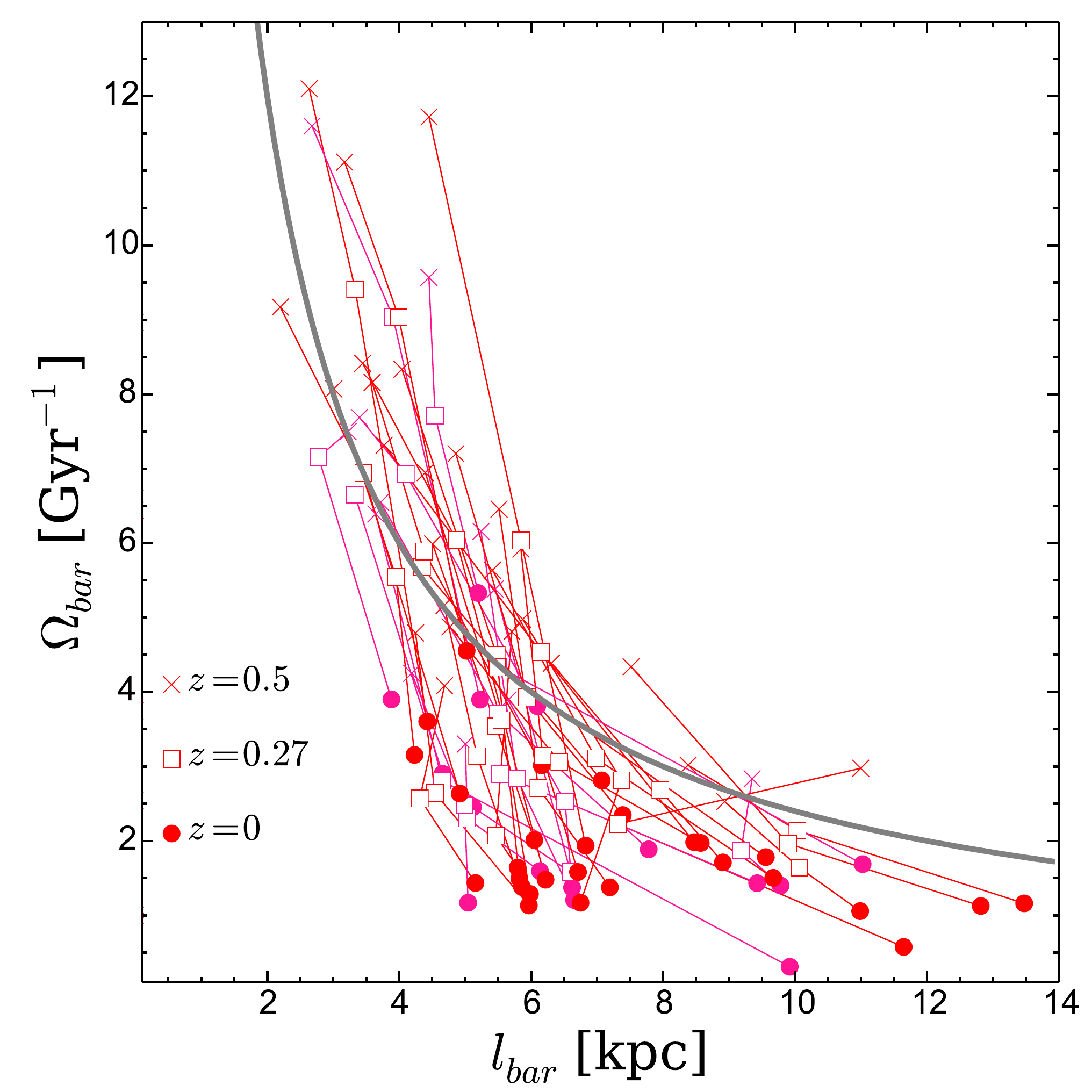}
\end{center}
\caption{Pattern speed vs bar length for strong bars. This shows
  clearly that bars slow down as they grow. At early times this
  follows roughly  the $l_{\rm
    bar}\propto \Omega_{\rm bar}^{-1}$ scaling expected for corotation
  radii in discs with flat rotation curves (grey dotted curve). At
  late times the pattern speed slows down with little further increase
  in bar length, pushing corotation well beyond the edge of the bar.
}
  \label{FigRbarOmega}
\end{figure} 
%%%%%%%%%%%%%%%%%%%%%%%%%%%%%%%%%%%%%%%%%%%%%%%%%%%%%%%%%%%%%%%

Fig.~\ref{FigFdiscFdec} thus suggests that combining both
$f_{\rm disc}$ and $f_{\rm dec}$ improves matters. For example, the
combined criteria $f_{\rm disc}<1$ and $f_{\rm dec}>0.95$ identify
$82\%$ of strong bars. Of galaxies satisfying these criteria, only
$11\%$ remain unbarred.  

Similarly, the criteria $f_{\rm disc}>0.95$ and $f_{\rm dec}<1$ single
out $77\%$ of all discs that remain unbarred. Of these, very few have
developed strong bars (only $10$) and $28$ out of a total of $59$ have
developed weak bars. The latter are, perhaps unsurprisingly, much harder
to predict on the basis of $f_{\rm disc}$ and $f_{\rm dec}$ alone, and
are seen to span nearly the full range of allowed values in
Fig.~\ref{FigFdiscFdec}.

We conclude that, in order to develop strong bars, discs must be
locally and globally dominant; in other words, they must contribute a
large fraction of the inner mass budget to systems where the disc
circular speed exceeds that of its halo. On the other
hand, galaxies that remain unbarred are predominantly those where the
disc is less important, not only within their half-mass radii but also
in relation to their surrounding halos.

\subsection{Bar slowdown}
\label{SecBarS}

As discussed in Sec.~\ref{SecIntro}, we expect bars that grow
gradually to slow down as they become stronger. This is indeed the
case in our simulations, as shown in Fig.~\ref{FigRbarOmega}. This
figure shows the decline in the bar pattern speed, $\Omega_{\rm bar}$,
since $z=0.5$ as a function of bar length. We only show this for the
``strong'' bars in our sample because of difficulties estimating
 the pattern speed of weak bars accurately.

Fig.~\ref{FigRbarOmega} shows that $\Omega_{\rm bar}$ has decreased on
average by a factor of $3$ over the past $5$ Gyrs. In the same time
interval bar lengths have increased by a factor of $1.7$ on
average. Indeed, the slowdown seems to roughly satisfy the
$l_{\rm bar}\times \Omega_{\rm bar}=$constant relation (grey dotted
line) expected for bar lengths that increase in proportion to the
corotation radius in a galaxy with a flat circular velocity profile. 

\begin{figure}
\begin{center}
\includegraphics[,width=1.0\linewidth,clip]{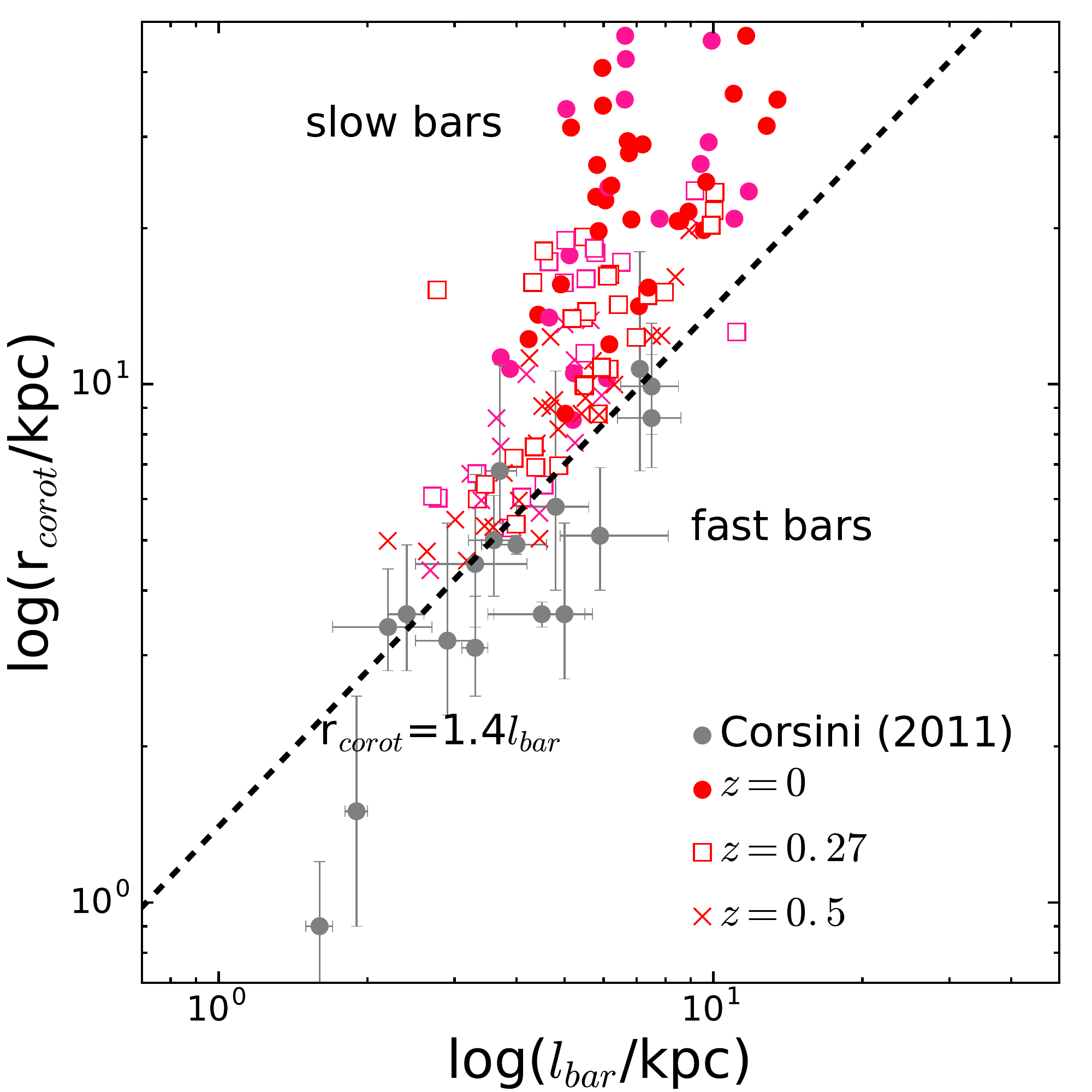}
\end{center}
\caption{Corotation radius vs bar length for {\it strong} bars in our
  sample, at three different times: $z=0.5$ (triangles), $z=0.27$
  (squares),  and $z=0$ (circles). Grey symbols with error bars are
  observational data from the compilation of
  \citet{Corsini2011}. ``Fast bars'' are those below the dotted line
  delineating $r_{\rm corot}=1.4\, l_{\rm bar}$. Most strong bars in our
  simulation are ``slow'' at $z=0$, in contrast with observational estimates.
  }
    \label{FigRbarRcor}
\end{figure}

At late times, the slowdown proceeds in
most galaxies without the corresponding increase in bar length, so
that bar lengths become smaller than their corotation radii
at $z=0$. We examine this in more detail in Fig.~\ref{FigRbarRcor},
where we show $r_{\rm corot}$ vs $l_{\rm bar}$ for all strongly barred
galaxies (as identified at $z=0$) at three different redshifts
($z=0.5$, $z=0.27$, and $z=0$) and compare them with the
compilation of \citet{Corsini2011}, which only includes galaxies in the
local universe.

Bars below the dotted line that delineates
$r_{\rm corot}=1.4\, l_{\rm bar}$ are usually referred to as ``fast
bars'', a characterization that describes well the few galaxies for
which pattern speeds have been reliably measured observationally. Note
that EAGLE strong bars, although relatively fast at early times
according to this characterization, have clearly become
slow\footnote{The few weak bars we were able to measure reliable
  pattern speeds for at $z=0$ are slightly faster, but still not as
  fast as observed. We do not include them in Fig.~\ref{FigRbarRcor}
  because we were unable to measure pattern speeds for all
  of them.} by $z=0$ confidently.

 %%%%%%%%%%%%%%%%%%%%%%%%%%%%%%%%%%%%%
%%%%%%%%%%%%%%%%%%%%%%%%%%%%%%%%%%%%% 
  
This result is reminiscent of the arguments of \citet{Debattista2000},
who argued that ``fast bars'' presented a challenge to $\Lambda$CDM
models. In their argument, dynamical friction in halos as centrally
concentrated as those expected in $\Lambda$CDM would quickly slow down
a bar and push its corotation radius well beyond the edge of the
bar, just as seen at $z=0$ in Fig.~\ref{FigRbarRcor}. 

The \citet{Debattista2000} observation ignited a spirited debate about
the true slowdown rate of bars in N-body simulations that is, as far
as we can tell, still unresolved \citep[see, e.g.,][and references
therein]{Sellwood2006,Weinberg2007a,Weinberg2007b,Sellwood2008}. The
disagreement centres on the role of dynamical friction vs resonances,
and on the minimum numerical resolution required to properly follow
the bar slowdown. In particular, \citet{Weinberg2007a} argue that
several {\it millions} of particles would be needed to capture the
resonant coupling that slows down the bar, and warn that slowdown
timescales may be severely underestimated in simulations with limited
numbers of particles.

%%%%%%%%%%%%%%%%%%%%%%%%%%%%%%%%%%%%%
\begin{figure}
\begin{center}
\includegraphics[,width=1.0\linewidth,clip]{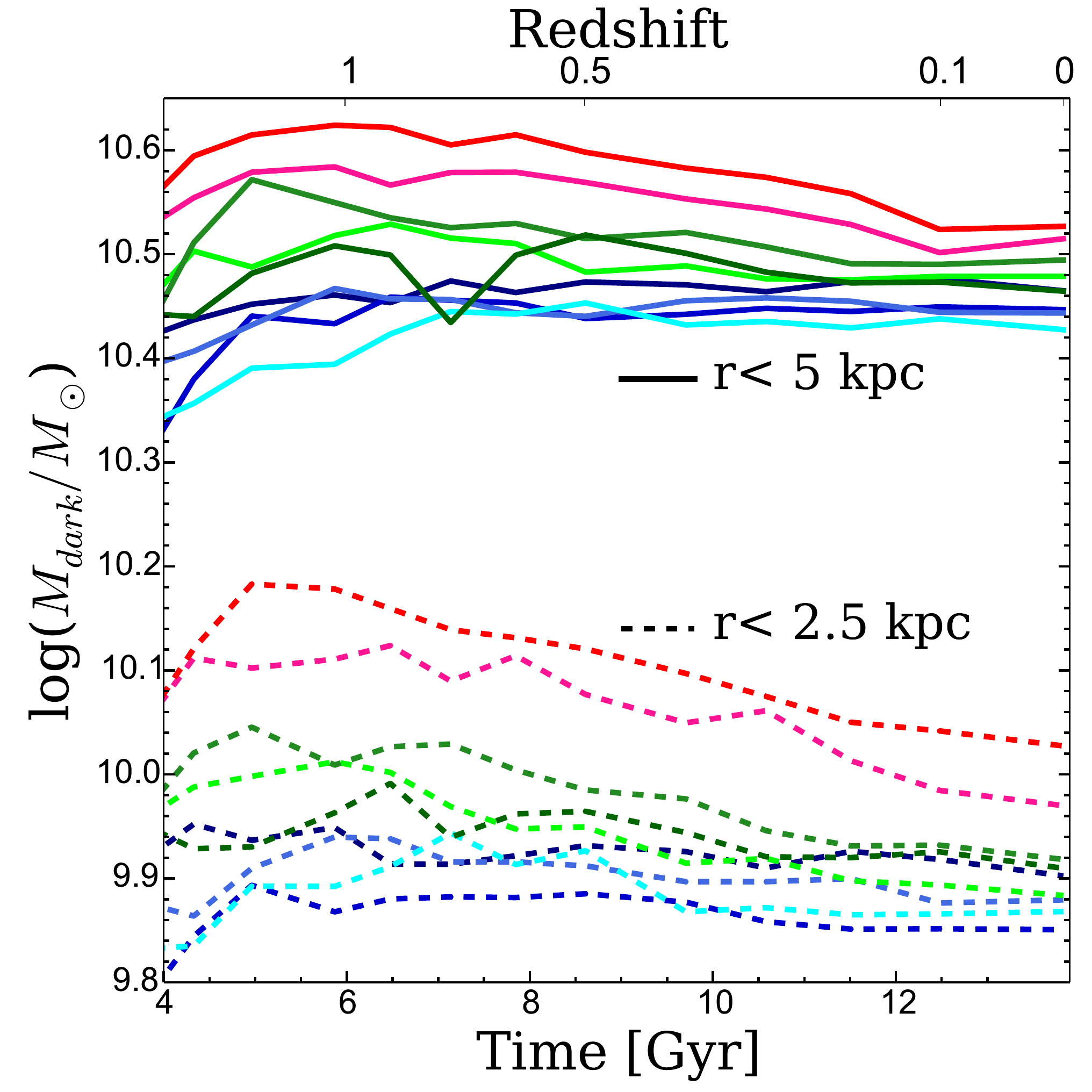}
\end{center}
\caption{Evolution of the dark matter mass enclosed within two fixed
  physical radii, $r=5$ kpc and $r=2.5$ kpc, averaged for galaxies
  binned as a function of their bar strength at $z=0$ (i.e., as in
  Fig.~\ref{FigBarG}). Bar slowdown clearly reduces the central
  density of dark matter within the region of the bar. By contrast,
  the central dark matter densities of unbarred galaxies remain
  unchanged during the past $5$-$6$ Gyr.
}
\label{FigEvMhalo}
\end{figure}
%%%%%%%%%%%%%%%%%%%%%%%%%%%%%%%%%%%%%

The numerical resolution of our simulations is admittedly poor by comparison, so it
is unclear whether our results may help to resolve this
disagreement. We therefore just note that strong bars slow down very
rapidly in our simulations, creating a population of ``slow bars''
that, apparently, have no obvious observational counterparts. If confirmed by
simulations with improved numerical resolution, this may very well
present a challenge to models of barred galaxy formation in a
$\Lambda$CDM universe. Resolving this issue, however, may require much
higher resolution simulations than achievable today for cosmologically
significant volumes.

\subsection{Halo evolution}
\label{SecHaloEvol}

The dramatic bar slowdown discussed in the previous section suggests
that the bar interacts strongly with the dark matter halo. This
interaction leads to substantial exchange of energy and angular
momentum, leading to substantial expansion of the inner regions of the
halo. We show this in Fig.~\ref{FigEvMhalo}, where we plot, for
various values of the bar strength, the evolution of the enclosed dark
matter mass within two different radii, $r=2.5$ and $5$ kpc. It is
clear from this figure that the slowdown of the bar induces
significant expansion of the inner dark matter mass
profile. 

Interestingly, even weak bars are able to lower the
central density of dark matter significantly, implying that non-axisymmetric
features in disc galaxies might be an important driver of the
transformation of the inner mass profiles of their dark matter
halos. By contrast, the inner regions of halos of disc galaxies that
do not develop a bar evolve little over the past $\sim 5$ Gyr or so.  

Finally, Fig.~\ref{FigEvMhalo} shows that, despite the bar-induced
expansion of the inner halo, barred galaxies at $z=0$ still have more
dark matter in their inner regions than their unbarred counterparts of
similar stellar and virial mass. This is mainly because barred
galaxies occur predominantly in dense, dominant discs that have
substantially contracted the dark matter profile before the bar
forms. 

Indeed, galaxies that will become strong bars have, at $z\sim 0.5$, roughly
twice as much dark matter within $2.5$ kpc from the centre as galaxies
that remain unbarred. Although the difference narrows as the bar slows
down, it still remains at $z=0$.  This implies that the bar-induced
halo expansion might not actually create a constant density ``core''
in a cuspy dark halo; the inner cusps of halos of would-be bars are so
steep because of contraction that even after expanding they may remain
cuspy. Although qualitatively our argument seems robust we are unable
to examine it quantitatively because of limited numerical resolution:
the average \citet{Power2003} convergence radius of our systems, for
example, is of order $\sim 5$ kpc. Clarifying whether bars carve
``cores'' out of cusps or not will need to await simulations with much
higher numerical resolution.

\section{Summary and Conclusions}
\label{SecConc}

We have used a $\Lambda$CDM cosmological hydrodynamical simulation
from the EAGLE project to study the formation of barred galaxies. The
simulation evolves a box $100$ Mpc on a side with $2 \times 1504^3$
particles, half of which are baryonic and half dark matter. Our study
focusses on a narrow range of stellar mass,
$10.6<\log{M_{*}/M_{\odot}} <11$, which are resolved with at least
22,000 star particles. Of the $495$ galaxies in that mass
range identified at $z=0$ we select a sample of $269$ ``discs'',
defined as flattened systems with minor-to-major axis ratio
$c/a < 0.4$ and relatively low vertical velocity dispersion
$\sigma_z/\sigma_{\rm tot}<0.5$.

We identify barred galaxies by measuring the amplitude of the
normalized $m=2$ Fourier mode of the azimuthal surface density profile
as a function of cylindrical radius, and choose as a measure of bar
strength the peak amplitude, $A_2^{\rm max}$. We consider ``barred''
all discs with $A_2^{\rm max}>0.2$. We follow the evolution of all
these galaxies in order to estimate bar growth timescales, to identify
which parameters predict the development of bars best, and to measure
the evolution of bar strength, length, and pattern speed.

Our main conclusions may be summarized as follows.
\begin{itemize}

\item About $40\%$ of EAGLE discs in our sample are barred, $20\%$ of them strong
  bars ($A_2^{\rm max}>0.4$) and another $20\%$ weak bars
  ($0.2<A_2^{\rm max}<0.4$). This bar frequency seems in reasonable
  agreement with observational estimates from
  \citet{Sheth2008,Barazza2008,Nair2010}.

\item Bars in our simulated discs span a wide range in terms of
  length. They typically extend beyond the stellar half-mass radius but
  rarely exceed the radius containing $90\%$ of the stellar mass, in good
  agreement with observational estimates. In terms of each of those
  radii, the median bar length and interquartile range is given by
  $l_{\rm bar}/r_{50}=1.53^{+0.42}_{-0.41}$ and
  $l_{\rm bar}/r_{90}=0.35^{+0.18}_{-0.10}$.

\item At $z=0$, bar strength correlates strongly with stellar
  half-mass radius (stronger bars form in smaller discs), hinting
  that, as expected from earlier work, bars develop preferentially in
  systems where the disc is gravitationally important. We also
  find that stronger bars develop in systems that are less gas-rich,
  and that have formed the bulk of their stars earlier than unbarred
  discs.

\item Strong bars in our sample develop relatively quickly before
  saturating over a few Gyrs. Weak bars are still growing in strength
  at $z=0$, and take much longer to develop, with characteristic
  timescales approaching or even exceeding a Hubble time. Even our
  strongest, fastest growing bars take roughly $4$-$5$ Gyr ( a few
  dozen disc rotations) to form fully.

\item The gravitational importance of the disc at its half-mass radius
  may be used to predict which galaxies will develop bars, but its
  predictive power may be enhanced by considering the overall
  importance of the disc in the system as a whole. Strong bars form in
  discs where baryons dominate and whose rotation speeds exceed the
  maximum circular velocity of the halo. Unbarred galaxies are discs
  where baryons are less important and whose rotation curves tend to
  rise in the outskirts.

\item Strong bars slow down quickly as they grow and, at $z=0$ are
   in the ``slow bar'' regime, $r_{\rm corot}/l_{\rm bar}>1.4$. This
  is in contrast with the few bars whose pattern speeds have been
  inferred observationally, all of which are ``fast''. This
  discrepancy may either imply that bar slowdown rates are
  artificially high in simulations at EAGLE resolution
  \citep[e.g.,][]{Weinberg2007a}, or, as argued in earlier work, that
  producing long-lasting ``fast bars'' is a real challenge for
  $\Lambda$CDM \citep[e.g.,][]{Debattista2000}.

\item The bar slowdown induces an expansion of the inner regions of
  the dark matter halo, as they capture the angular momentum of the
  forming bar. However, bars form in massive dense discs with heavily
  contracted halos, so despite the bar-induced expansion barred
  galaxy halos are still more centrally concentrated than unbarred
  galaxies of similar stellar mass. Our numerical resolution is not
  enough to let us ascertain whether this expansion may lead to the
  formation of constant density ``cores'' in barred galaxy halos.

\end{itemize}

Our overall conclusion is that current $\Lambda$CDM cosmological
hydrodynamical simulations of cosmologically significant volumes such
as EAGLE yield a population of simulated discs with bar fractions,
lengths, and evolution that are in broad agreement with observational
constraints. They also confirm earlier suggestions that ``slow
bars'' might pose a severe challenge to this scenario. Although bars
form in the manner and frequency expected, they slow down too fast
through interaction with the dark halo. Unless the Universe has a
population of slow bars that has yet to be recognized, or the bar
slowdown we measure is artificially enhanced by limited numerical
resolution, accounting for the presence of ``fast bars'' in
strongly-barred discs is a clear goal for the next generation of
$\Lambda$CDM simulations of galaxy formation.

\section*{ACKNOWLEDGEMENT}

DA, MGA, JFN, LVS acknowledge financial support from grant
PICT-1137/2012 from the Agencia Nacional de Promoci\'on Cient\'ifica y
Tecnol\'ogica, Argentina. MGA acknowledges financial support of grant
203/14 of the SECYTUNC, Argentina. We are grateful to Alejandro
Ben\'itez-Llambay for the use of his Py-SPHViewer software. The
research was supported in part by the European Research Council under
the European Union´s Seventh Framework Programme (FP7/2007-2013) / ERC
Grant agreement 278594-GasAroundGalaxies, 267291- COSMIWAY, by the
Interuniversity Attraction Poles Programme initiated by the Belgian
Science Policy Office (AP P7/08 CHARM) and by the Netherlands Organisation for Scientic 
Research (NWO), through VICI grant 639.043.409. This work used the DiRAC Data
Centric system at Durham University, operated by the Institute for
Computational Cosmology on behalf of the STFC DiRAC HPC Facility
(www.dirac.ac.uk). This equipment was funded by BIS National
E-infrastructure capital grant ST/K00042X/1, STFC capital grants
ST/H008519/1 and ST/K00087X/1, STFC DiRAC Operations grant
ST/K003267/1 and Durham University. DiRAC is part of the National
E-Infrastructure. RAC is a Royal Society University Research Fellow.
CDV acknowledges financial support from the Spanish Ministry of Economy and 
Competitiveness (MINECO) under the 2015 Severo Ochoa Program 
SEV-2015-0548 and grant AYA2014-58308.

\bibliographystyle{mnras}
\bibliography{biblio}

\bsp	% typesetting comment
\label{lastpage}

\end{document}